Consensus, Bi-polarization and Multiformity in Opinion Dynamics with Bidirectional Thresholds

Shuo Liu[a], Xiwang Guan[a], Shuangling Luo[b], Haoxiang Xia[a*]

[a] Institute of Systems Engineering, Dalian University of Technology, Dalian 116024, China
[b] School of Maritime Economics and Management, Dalian Maritime University, 116026, China

**Abstract**.
Many empirical networks are intrinsically pluralistic, with interactions occurring within groups of arbitrary agents. Then the agent in the network can be influenced by types of neighbors, common examples include similarity, opposition, and negligibility. Although the influence of neighbors can be described as an amicable and antagonistic relationship in complex real-world systems accurately, and the research on the dynamic process of public opinion evolution with different types of influence is valuable, few studies have mentioned that issue. In this paper, we develop a novel model on networks of agents with the bi-directional bounded thresholds for studying the evolution of opinion dynamics. We define the scope of individual assimilation and exclusion to identify different types of neighbors and calculate the impact of the corresponding neighbors on the individuals by converting the opinion difference. The simulation results show that the proposed mechanism can effectively explain the formation of bi-polarization during opinion evolution and the settings of the bi-directional bounded thresholds significantly influence the eventual distribution of opinions. Furthermore, we explore the impacts of the initial conditions and the structure of the small-world network on the evolution of opinions.

## 1. Introduction

The evolution of opinions within a population through the interaction between individuals, as a macroscopic collective social phenomenon, has received increasing attention in social psychology, anthropology, physics, economics, and mathematics[1-2]. The analysis of the underlying mechanism of opinion dynamics can help understand the dynamics of, e.g., collective decision making, shifts in preferences, minority opinion survival, rise of extremism, political changes, the emergence of fads and the like[3-6].

The early research on group behavior by sociologists and social psychologists has provided many valuable ideas and methods for further exploring the mechanism of public opinion[7-9]. In recent years, with the intersection and integration of different disciplines, scholars have established a series of dynamic models by means of mathematical and physical methods for exploring the formation mechanism of public opinion. The Sznajd model was proposed by Sznajd-Weron and Sznajd in 2000 is one of the most popular models on public opinion, which is a relatively simple model and based on the Ising model[10]. Furthermore, Staufer et al discussed a more general Sznajd model and mapped the Galam model to a one-dimensional Sznajd model[11-12]. In addition, the opinion dynamics model also includes the voter model[13], and the majority-rule model[14] Deffuant and Weisbuch model (DW model)[15] and Hegselmann and Krause model(HK model)[16-17].

*E-mail address: hxxia@dlut.edu.cn

Agent-based models are one of the most powerful tools available for theorizing about opinion dynamics[18-20]. In an agent-based model of opinion dynamics, each agent has an opinion which might be described by an either continuous or discrete variable. Relationships between agents, such as their family relation and acquaintance, are usually represented utilizing a social network. Agents can influence each others' opinions through the connections existing between them by exercising some common rule of opinion updating. Among these models, the model of continuous opinion dynamics under bounded confidence has received significant attention. Reflecting the well-known aphorism "Like likes like", in the bounded confident model(BCM), agents only interact with each other when their opinions are very similar[21]. However, the bounded-confidence assumption has also been subject to important criticism: the model assumes that individuals always ignore opinions that differ from their views. But in real word, people are not only influenced by their friends or similar agents positively but also negatively influenced by their enemies' opinion[22-23] or dissimilar agents[24].

To describe that dynamic process, the mechanism of negative impact has been proposed in a few opinion dynamics researches. The mechanisms used in the literature[25-27] are all based on the DW model, which indicates that only one pair of individuals in the group is selected at each time step to interact with each other. This is not enough for us to be affected by many aspects at the same time in the era of information explosion. In this work, we extend the HK model to study the impact of both positive and negative effects on the evolution of public opinion on complex networks. We set two bounded thresholds to define the space of positive and negative impacts, and then model the dynamics of individual opinions on the network, observe the impact of different threshold characteristics on the evolutionary results of public opinion, and analyze the impact of initial conditions. Compared with previous works, our main contribution is that we consider the influence of different types of neighbors on individual opinions, and the impact of initial biased opinion groups and different network structures on the evolution of opinions.

The remainder of the paper is organized as follows. In Section 2, we describe the continuous opinion model with the bi-directional bounded threshold (BBT). Our main results are presented in Section 3, devoted to the study of the opinion dynamics according to BBT to different threshold values and initial conditions by numerous simulations. Finally, the summary and conclusions are given in Section 4.

## 2. Model

The HK model[17] is a simple model which describes multiple neighbors satisfying the conditions simultaneously affect individual opinions. Suppose that there are N agents in a social network, each agent has an opinion $x_i(t) \in R$ at a discrete time $t$. The opinion $x_i(t)$ varies continuously within an opinion interval. In the HK model, agent $i$ updates his opinion by averaging all opinions of the neighbors whose opinions lie in the confidence range of agent $i$, that is, the opinion of agent $i$ is updated by

$$x_i(t+1) = \frac{1}{|A_i(t)|} \sum_{j \in A_i(t)} x_j(t), i = 1, \cdots, N; t = 0,1,2 \cdots, \qquad (1)$$

Where $A_i(t) = \{j \mid |x_i(t) - x_j(t)| < \varepsilon\}$ denotes the index set of the neighbors of agent $i$ and $|A_i(t)|$ is the cardinality of the neighbor set. As shown in Fig1(a), individuals are only affected by neighbors with similar opinions, while others are ignored.

By extending this HK model, we in this work use an additional threshold to depict the negative communications, through which dissimilar opinions become even more dissimilar. In the classical HK model, individuals only consider all the neighbors that are similar to themselves, and replace their opinions with the average of the opinions of all similar neighbors at the next step. However, two bounded thresholds are involved in this model, namely the positive influence threshold and the negative influence threshold, respectively, to identify neighbors with similar opinions and neighbors with large differences as shown in Fig1(b). In this model, individuals update their opinions not only by considering all neighbors that are similar to themselves, but also by considering the impact of their own opinions at the previous time step and all neighbors with great differences from their own opinions.

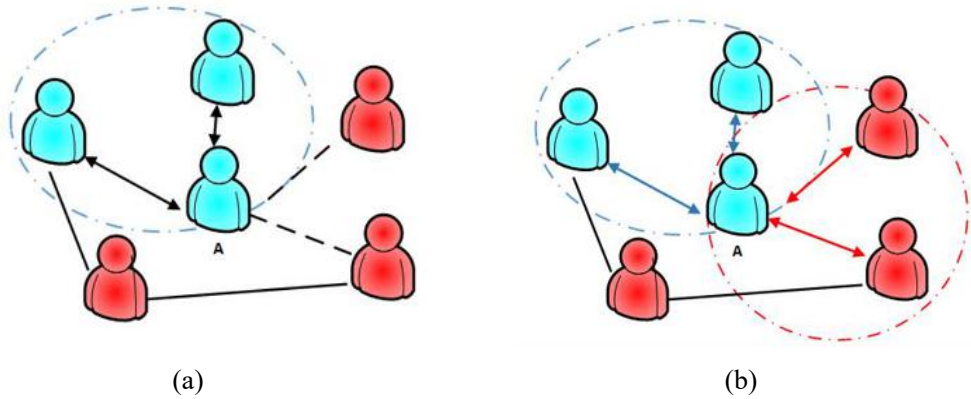

(a)            (b)

Fig.1. Illustration of the proposed model. The solid edges denote connecting influential individuals and the dashed edges represent links which connect no effect neighbors. Blue individuals represent neighbors that have a positive impact on individual A, and red indicates a neighbor with negative influence. In (a) the original HK model is illustrated, in which individuals are only influenced by individuals whose opinions are similar to their own, ignoring all other situations. In (b) the model suggested in this work is illustrated, in which individuals are affected not only by individuals with similar opinions but also by individuals with large differences in opinions.

The proposed model is more formally described in the following.

(i) We first define two thresholds, namely a threshold for positive influence $\varepsilon_p$ and a threshold for negative influence $\varepsilon_n$, in order to incorporate bi-directional opinion influence in the model. With these two thresholds, the neighborhood of agent $i$ can be categorized into three groups, i.e. the neighbors with positive influence, the neighbors with negative influence, and the others. The first two groups can respectively be defined by $A_p(i,x(t)) = \{j \| x_i(t) - x_j(t) \| < \varepsilon_p\}$ and $A_n(i,x(t)) = \{z \| x_z(t) - x_i(t) \| > \varepsilon_n\}$.

(ii) The average of the positive impact for agent $i$ is obtained by $\overline{x_j(t)} = \frac{1}{|A_p(i,x(t))|} \sum_{j \in A_p(i,x(t))} x_j(t)$, and the average of the negative impact for agent $i$ is

$$\overline{x_z(t)} = \frac{1}{|A_n(i,x(t))|} \sum_{z \in A_n(i,x(t))} x_z(t) .$$

(iii) We calculate how much agent $i$'s opinion is positively affected,

$$\Delta E_p = (\overline{x_j(t)} - x_i(t)) \tag{2}$$

And how much is negatively affected,

$$\Delta E_n = \begin{cases} \lambda * \dfrac{\varepsilon_p}{1-\varepsilon_n} (|\overline{x_z(t)} - x_i(t)| - \varepsilon_n) & , \varepsilon_p \neq 0 \\ \lambda * (\overline{x_z(t)} - x_i(t)) & , \varepsilon_p = 0 \end{cases} \tag{3}$$

where $\lambda = -\dfrac{(\overline{x_z(t)} - x_i(t))}{|(\overline{x_z(t)} - x_i(t))|}$ is directional coefficient which describes the opposite direction of negative impact opinions.

(iv) Then the agent update its opinion at time $t$ by:

$$x_i(t+1) = B[x_i(t) + \gamma * \Delta E_p * \mathrm{sgn}(\varepsilon_p) + \gamma * \Delta E_n * \mathrm{sgn}(1-\varepsilon_n)] \tag{4}$$

where $\mathrm{sgn}(:)$ is a signal function; $\gamma$ is the coefficient of external influence; $B[x]$ is a limit function, specifically described as follows,

$$B[x] = \begin{cases} 1 & x > 1 \\ x & 0 \leq x \leq 1 \\ 0 & x < 0 \end{cases} \tag{5}$$

The preceding process is to be repeated until the system evolves to a steady state, where the nodes states in the network no longer change $x_i(t+1) - x_i(t) \leq 10^{-4}, i \in N$ and the system reaches the dynamic equilibrium. This is the whole process of non-Monte-Carlo simulation on networks. It should be noted that the structure of the network does not change during the whole evolutionary process.

## 3. Results

In order to observe the function of different degrees of both positive and negative effects, in this section, we apply the model to complete networks with different bi-directional bounded thresholds. After that, the model is also used to explore the impact of biased initial opinion values and different structure of complex.

### 3.1 Clarifying the influence of bi-directional bounded threshold

We performed comparative experiments with different bi-directional bounded threshold settings using complete networks with randomly distribution of initial opinion of nodes. Therefore, the impact of different positive and negative influence thresholds on evolutionary results can be observed. In the simulation, we set $N = 200$, $c = 0.1$, $\varepsilon_n \geq \varepsilon_p$, and carry out public opinion evolution according to the model described in Section 2. We generated 50 networks, and each network evolves for 1000 times, so as to avoid the effect of contingency on experimental results.

The simulation results are shown in Fig.2(a) and Fig.2(b) which represent the number of opinion clusters and the maximum opinion difference in the network respectively, when the network reach a steady state. The heat maps show the evolutionary results under different parameter sets, plotted by positive influence threshold on the horizontal axis and negative influence threshold on the vertical, and the color represents the number of opinion clusters.

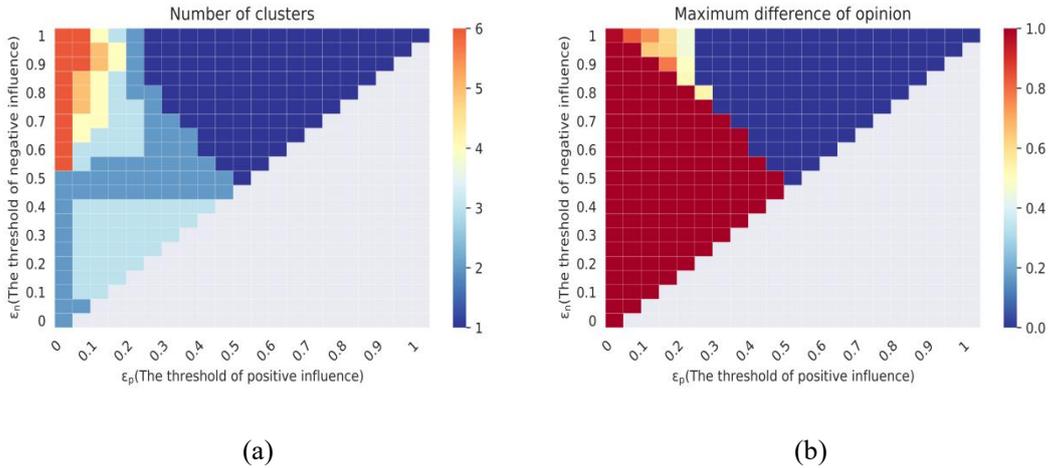

(a)          (b)

Fig.2. The evolutionary results on the complete network with randomly distribution of initial opinions, (a) represents the opinion cluster in the network when the network reaches stability, while (b) represents the maximum opinion difference in the network when the network reaches stability.

According to the results shown in Fig.2(a) and Fig.2(b), the following conclusions are obtained:

(i) When the threshold of negative influence is 1, the opinion update in the group will not be affected by the negative opinion, and the evolution result of group opinion is consistent with the classical HK model[17]. It can be seen in Fig.2 (a) that when the threshold of positive influence is greater than 0.25, the group opinion reaches consensus. The smaller the threshold of positive influence is, the larger the number of opinion clusters in the group will be. When the threshold of positive influence is 0.2, it can be concluded from Fig.2 (b) that the difference between the two groups' opinions is about 0.5, which is called polarization in the literature, but the two groups' opinions are not extreme opinion groups.

(ii) Based on this new model(BBT), we get that when the threshold of negative influence is

greater than 0.5, the consensus of the opinions will occur in the group; when the threshold of the negative influence is gradually increased, the threshold of the positive influence required for the consensus of the group is gradually reduced. When the threshold of positive impact is greater than 0.5, the group will certainly form a consensus of opinion, as shown in Fig.3(a).

(iii) We can see that the dark red area in Fig.2(b) which indicates that there are two extreme opposing groups in the group when the evolution of opinions goes steady. In the area, where the positive impact is 0 (indicating that the opinions of any individual in the group do not converge with each other), the negative opinion is less than 0.5 and the negative impact threshold increases from 0.45 to 0.75 as the positive impact threshold varies between 0 and 0.45, the group forms two extreme opposite parts, as shown in Fig.3(b). In this paper, we clarify that the bi-polarization is the situation that the opinion of the group divide two parts and the difference between them become increasingly larger till completely opposite according to the interaction of individual opinions.

(iv) When the ranges are the negative influence threshold and positive influence threshold are both small in the range of [0,0.45] and the positive influence threshold is between 0.1 and 0.25, the negative influence threshold is between 0.6 and 0.75, there are three clusters in the final opinion group, reflecting the situation where two extreme opinion clusters and one intermediate opinion cluster coexist. However, the mechanism of the three clusters formed by these two regions is completely different. When the two thresholds are small, the intermediate opinion cluster is subject to the common rejection of the upper and lower extreme groups and the repulsive force is quite concentrated in the middle opinion. When the positive impact threshold is small, the negative impact threshold is larger, and the formation of the three opinion clusters It is because the middle opinion cluster is not attracted or excluded by the upper and lower extreme opinion clusters, and is concentrated in the middle, as shown in Fig.3(c). And we also find that when the positive influence threshold is between 0 and 0.15, the number of group opinion clusters increases with the negative influence threshold gradually increasing from 0.6 to 1 and the number of non-extreme groups has increased. When the negative impact threshold increases to around 0.9, with the increase of the positive impact threshold, the two extreme groups are weakened by the extreme opposition groups, and the positive influence of the neighbor groups is enhanced, resulting in the disappearance of the two extreme opinion clusters, as shown in Fig.3(d).

(v) Based on the analysis of the results of the two graphs, the evolution results of our model can be divided into the following four types: Consensus; two-extreme-opposites; the coexistence of two extreme groups an neutral opinion clusters; and the coexistence of multiple opinion clusters in the middle, as shown in Fig.3(a-d).

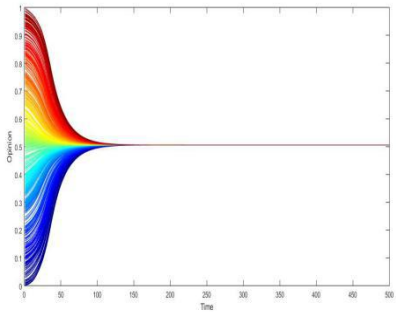 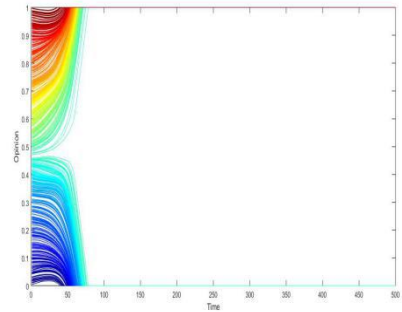

(a)            (b)

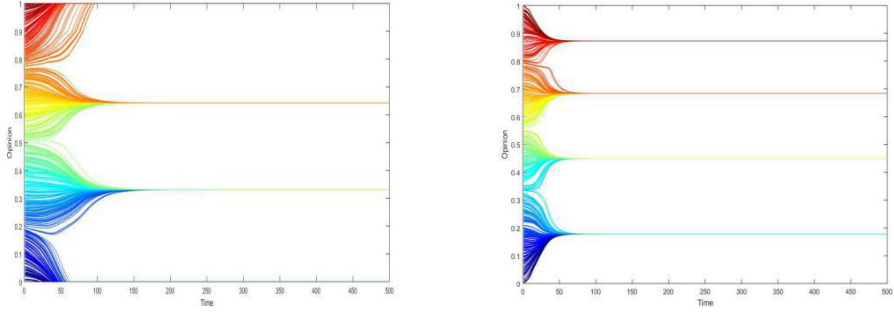

(c)　　　　　　　　　　　　　(d)

Fig.3. The processes of node's opinion evolution in the network under different bi-di-rectional bounded threshold settings (a) $\varepsilon_p = 0.6, \varepsilon_n = 0.7$ .(b) $\varepsilon_p = 0.4, \varepsilon_n = 0.5$ .(c) $\varepsilon_p = 0.1, \varepsilon_n = 0.7$ . (d) $\varepsilon_p = 0.2, \varepsilon_n = 0.9$ .

### 3.2 The influence of biased initial opinion

In reality, when people facing different types of messages, the original feedback of their attitude is different. When receiving scientific information, most people in the group tend to support opinions; when receiving controversial information, individuals will choose to support or disagree according to their own perceptions[28]. Therefore, it is necessary to explore the tendency of the initial opinions generated by the group in the face of different information[29].

In this section, we apply our model to the groups with different degrees of initial opinion bias. In order to explore the influence of the biased initial opinion, we choose complete network to ignore the impact of the network structure on the evolution of opinions. The remaining parameters are consistent with the previous section. 1 stands for is fully supporting, while 0 stands for completely opposing. The simulation results of bi-directional bounded thresholds under the condition of initial opinions with different degrees of tendentiousness are given in Fig.4, which show the number of final opinion clusters and the maximum opinion difference of the network in the steady state when the propensity proportions of the supporters at the initial time are 0%, 10%, 20%, 30%, and 40%, respectively.

Combined with the two sets of graphs with the final number of opinion clusters and the maximum difference of opinion in different bi-directional thresholds, we can find that when there is no individual inclined to support the event in the group, the positive influence threshold required to reach consensus in the network is greatly reduced. When there are proponents in the network, the proportion does not have a significant impact on the conditions needed to reach consensus. When the number of people inclined to support and oppose is about equal, the network is more likely to form multiple opinion clusters when the positive influence threshold is small.

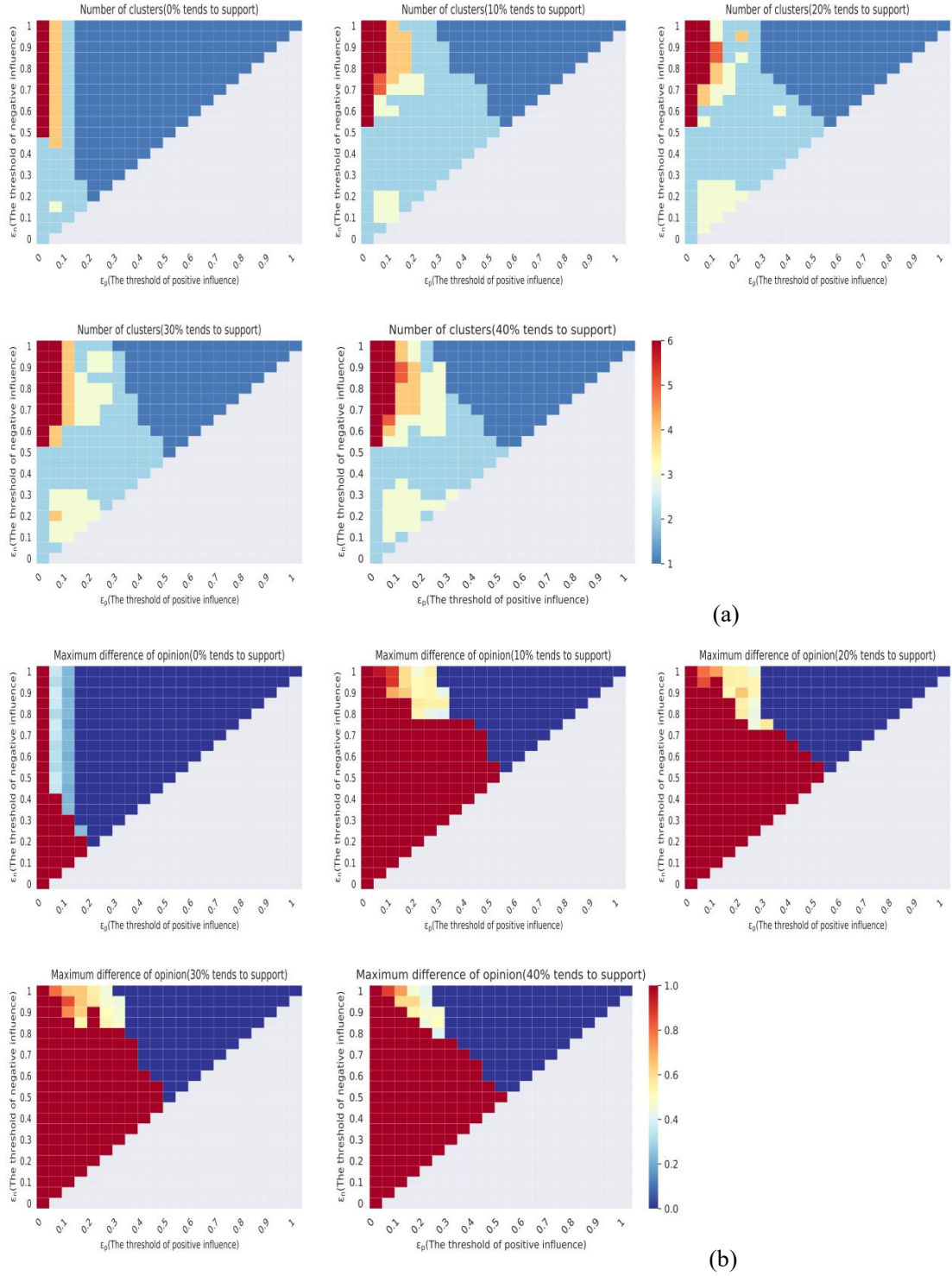

Fig.4. The evolutionary results on the complete network with biased initial opinions(0%,10%,20%,30%,40% trends to support), (a) represents the opinion cluster in the network when the network reaches stability (b) represents the maximum opinion difference in the network when the network reaches stability.

### 3.3 The influence of small-world network structure

Small-world network and scale-free network were proposed to effectively explain some structures in reality[30]. In this part, we mainly study the influence of the re-connection probability of the small-world network on the opinion evolution based on this model. We set the initial opinion random distribution, and $N = 10^3$, $k = <6>$, the re-connection probability respectively( $p$ ) selected by 0,0.0001,0.001,0.01,0.1. And the representative thresholds of positive influence and negative influence were chosen to be $\varepsilon_p = \varepsilon_n = 0.25$, $\varepsilon_p = \varepsilon_n = 0.5$, $\varepsilon_p = \varepsilon_n = 0.75$.

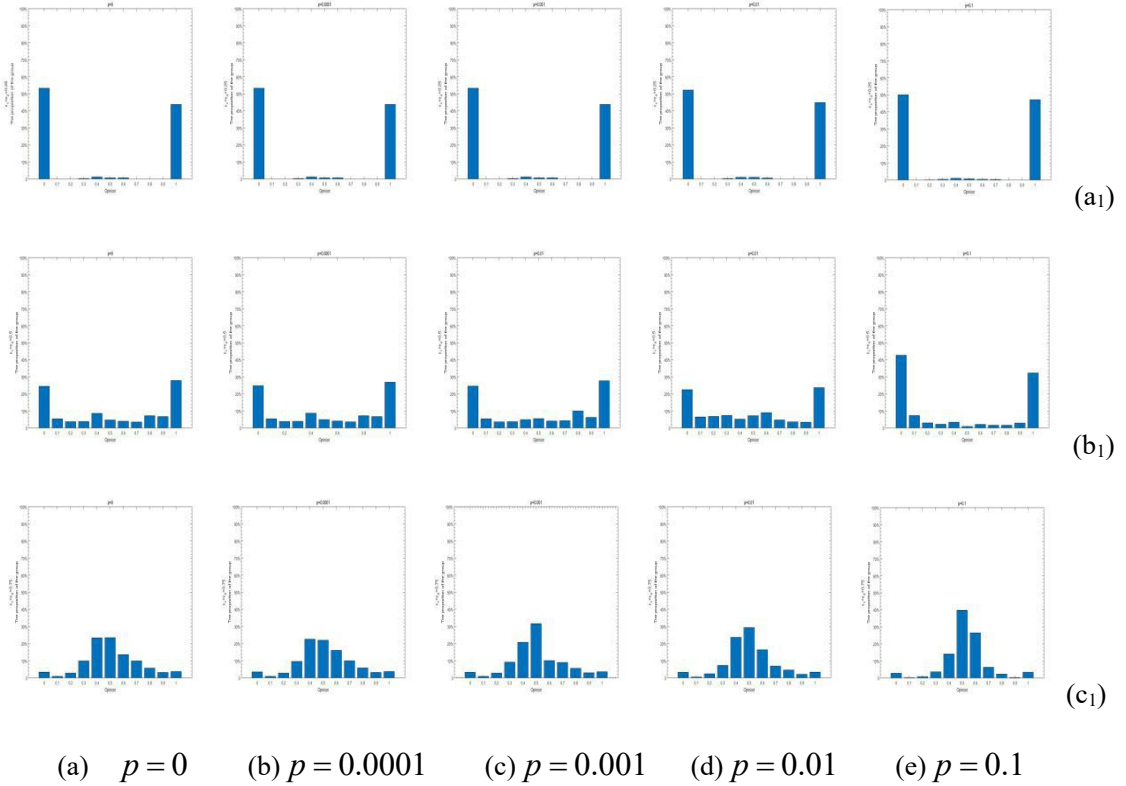

(a) $p = 0$ (b) $p = 0.0001$ (c) $p = 0.001$ (d) $p = 0.01$ (e) $p = 0.1$

Fig.5. The results on the small-world network with different re-connection probability $p = 0,0.0001,0.001,0.01,0.1$. ($a_1$)The thresholds of positive influence and negative influence are 0.25. ($b_1$)The thresholds of positive influence and negative influence are 0.5. ($c_1$)The thresholds of positive influence and negative influence are 0.75.

According to Fig.6., the results show the proportion of individuals in different opinion clusters when the opinion of the network reach stabilization. The figures are plotted by opinion value between 0 and 1 on the horizontal axis and the proportion on the verticals. From the figures we can see that when the thresholds of positive and negative influence are 0.25, the probability of re-connection has no significant effect on the evolution of opinions. When the thresholds of positive and negative influence are 0.5, the number of the two extreme opinion groups increases significantly until the probability of re-connection increased to 0.1. When the thresholds of positive and negative influence are 0.75, the number of individuals in the intermediate opinion cluster increases with the probability of re-connection increasing. Combining these results with the previous results, we found that even the thresholds of positive influence and negative influence reach 0.75, individuals holding extreme opinion still exist in the network. The increasing of the probability of re-connection promotes the interaction of individuals' opinions except for the case that thresholds of positive influence and negative influence are both small. Under the structure of the small-world network, although the threshold of positive influence and negative influence are as large as 0.75, there can still be individuals with extreme opinions in the network. This result can explain how the minority opinions' survival when the opposite opinion is predominating [5,31].

## 4. Conclusion

In this paper, we discuss continuous opinion dynamics models with the thresholds of positive and negative influence .In complex real-world systems, individual opinions receive both positive and negative influences from their neighbors and the public opinion evolves towards some direction to make the system more stable. And to model the dynamic process, we construct bi-directional threshold which is the threshold of positive influence and the threshold of negative influence. Individual update its opinion based on bi-directional thresholds. We observe the influence of different positive and negative influence thresholds on the evolution of network opinions, as well as the influence of initial opinions and network structure and pay attention to the re-connection probability of the small-world network.

Compared with the original opinion dynamics model, it can be seen that the addition of negative influence thresholds can well explain the phenomenon of increasing opinion differences in the population. Meanwhile, compared with the classical Deffuant–Weisbuch (DW) model and Hegselmann–Krause (HK) model, the positive influence threshold required for reaching consensus is greater than 0.25, and the negative influence threshold inhibits the formation of consensus. The more uniform the initial opinions are, the greater the possibility of more opinion clusters in the network when the positive influence threshold is small and the negative influence threshold is large. The connectivity of the small-world network promotes the interaction of opinions but the function is not evident when the positive and negative impact thresholds are small. Moreover, this model simulates both positive and negative effects on individual opinion updating. In the future, this model can be considered on different networks and Different research questions.